\documentclass[twocolumn]{aastex62}

\usepackage{amsmath}    
\usepackage{graphicx}   
\usepackage{verbatim}   
\usepackage{xspace}

\usepackage{amssymb}

\usepackage{ifthen}
\newboolean{isDraft}  \setboolean{isDraft}{false}

\usepackage{bm}
\usepackage{esvect}
\usepackage{gensymb}
\usepackage[ddmmyyyy]{datetime}

\usepackage[caption=false]{subfig}
\DeclareCaptionListOfFormat{myparens}{#2}
\newcommand{\Subref}[1]{\protect\subref{#1}}

\definecolor{white}{rgb}{1,1,1}
\definecolor{red}{rgb}{1,0,0}
\definecolor{blue}{rgb}{0,0,1}
\definecolor{darkBlue}{rgb}{0,0,.4}
\definecolor{green}{rgb}{0,1,0}
\definecolor{magenta}{rgb}{1,0,.6}
\definecolor{lightblue}{rgb}{0,.5,1}
\definecolor{lightpurple}{rgb}{.6,.4,1}
\definecolor{gold}{rgb}{.6,.5,0}
\definecolor{orange}{rgb}{1,0.4,0}
\definecolor{hotpink}{rgb}{1,0,0.5}
\definecolor{newcolor2}{rgb}{.5,.3,.5}
\definecolor{newcolor}{rgb}{0,.3,1}
\definecolor{newcolor3}{rgb}{1,0,.35}
\definecolor{darkgreen1}{rgb}{0, .35, 0}
\definecolor{darkgreen}{rgb}{0, .6, 0}
\definecolor{darkred}{rgb}{.75,0,0}
\xdefinecolor{olive}{cmyk}{0.64,0,0.95,0.4}
\xdefinecolor{purpleish}{cmyk}{0.75,0.75,0,0}
\definecolor{grayBold}{rgb}{0,0,0}
\definecolor{battleshipgrey}{rgb}{0.52, 0.52, 0.51}
\definecolor{cinereous}{rgb}{0.6, 0.51, 0.48}
\definecolor{darkgray}{rgb}{0.66, 0.66, 0.66}
\definecolor{davysgrey}{rgb}{0.33, 0.33, 0.33}
\definecolor{dimgray}{rgb}{0.41, 0.41, 0.41}
\definecolor{gainsboro}{rgb}{0.86, 0.86, 0.86}
\definecolor{grullo}{rgb}{0.66, 0.6, 0.53}
\definecolor{manatee}{rgb}{0.59, 0.6, 0.67}
\definecolor{oldlavender}{rgb}{0.47, 0.41, 0.47}
\definecolor{oldmauve}{rgb}{0.4, 0.19, 0.28}
\definecolor{khaki}{rgb}{0.76, 0.69, 0.57}
\definecolor{paynesgrey}{rgb}{0.25, 0.25, 0.28}
\definecolor{platinum}{rgb}{0.9, 0.89, 0.89}
\definecolor{whitesmoke}{rgb}{0.96, 0.96, 0.96}
\definecolor{coolblack}{rgb}{0.0, 0.18, 0.39}
\definecolor{royalblue}{rgb}{0.25, 0.41, 0.88}

\definecolor{red0}{RGB}{243,61,99}
\definecolor{red1}{RGB}{198,40,40}
\definecolor{green0}{RGB}{139,195,74}
\definecolor{blue0}{RGB}{33,150,243}
\definecolor{blue1}{RGB}{57,73,171}
\definecolor{blue2}{RGB}{134, 165, 216}
\definecolor{blue3}{RGB}{14,117,194}
\definecolor{blue4}{RGB}{10,86,143}

\definecolor{lightblack}{RGB}{55,59,61}

\definecolor{blueTbl}{RGB}{17, 143, 237}
\definecolor{greenTbl}{RGB}{137, 197, 65}

\definecolor{blueTblFaint}{RGB}{207, 232, 252}
\definecolor{greenTblFaint}{RGB}{231, 243, 216}

\definecolor{grayBckCol}{RGB}{244, 244, 244}
\definecolor{faintGrayBckCol}{RGB}{246, 246, 246}

\definecolor{figColBox}{rgb}{1,1,1}
\definecolor{linkCol}{RGB}{59,80,125}

\hypersetup{linkcolor=linkCol,citecolor=linkCol,filecolor=linkCol,urlcolor=linkCol,breaklinks=true}

\let\orgautoref\autoref
\providecommand{\Autoref}
{\def\equationautorefname{Equation}%
\def\figureautorefname{Figure}%
\def\subfigureautorefname{Figure}%
\def\chapterautorefname{Chapter}%
\def\sectionautorefname{Section}%
\def\subsectionautorefname{Section}%
\def\subsubsectionautorefname{Section}%
\def\Itemautorefname{Item}%
\def\tableautorefname{Table}%
\def\appendixautorefname{Appendix}%
\orgautoref}

\renewcommand{\autoref}
{\def\equationautorefname{Eq.}%
\def\figureautorefname{Fig.}%
\def\subfigureautorefname{Fig.}%
\def\chapterautorefname{Ch.}%
\def\sectionautorefname{Sect.}%
\def\subsectionautorefname{Sect.}%
\def\subsubsectionautorefname{Sect.}%
\def\Itemautorefname{item}%
\def\tableautorefname{Table}%
\orgautoref}

\newcommand{\ie}{i.e.\/,\xspace}

\newcommand{\eg}{e.g.\/,\xspace}
\newcommand{\Eg}{E.g.\/,\xspace}
\newcommand{\cf}{cf.\/\xspace}

\newcommand{\insitu}{{in-situ}\xspace}

\newcommand{\etc}{etc.\/\xspace}

\newcommand{\Sim}{\sim\kern-0.2em\xspace}

\newcommand{\mrm}[1]{\ensuremath{\mathrm{#1}}}
\newcommand{\txt}[1]{\text{#1}}
\newcommand{\tit}[1]{\textit{#1}}

\newcommand{\powA}[2]{\ensuremath{{{#1}\cdot10^{{#2}}}}\xspace}
\newcommand{\powB}[1]{\ensuremath{{10^{{#1}}}}\xspace}  

\newcommand{\gamlib}{\texttt{GammaLib}\xspace}
\newcommand{\ctools}{\texttt{ctools}\xspace}
\newcommand{\prodIrf}{\texttt{prod3b-v1}\xspace}
\newcommand{\tensorflow}{\texttt{tensorflow}\xspace}

\newcommand{\kev}{\ensuremath{\txt{keV}}\xspace}

\newcommand{\gev}{\ensuremath{\txt{GeV}}\xspace}
\newcommand{\tev}{\ensuremath{\txt{TeV}}\xspace}

\newcommand{\erg}{\ensuremath{\txt{erg}}\xspace}

\newcommand{\dgr}{\ensuremath{\degree}\xspace}

\newcommand{\hess}{{H.E.S.S.}\xspace}
\newcommand{\magic}{{MAGIC}\xspace}
\newcommand{\veritas}{{VERITAS}\xspace}
\newcommand{\fermi}{{\textit{Fermi}}\xspace}
\newcommand{\fermil}{{\textit{Fermi}-LAT}\xspace}
\newcommand{\fermig}{{\textit{Fermi}-GBM}\xspace}

\newcommand{\cta}{{CTA}\xspace}
\newcommand{\iact}{\txt{IACT}\xspace}
\newcommand{\iacts}{\txt{IACTs}\xspace}

\newcommand{\roi}{RoI\xspace}
\newcommand{\rois}{RoIs\xspace}
\newcommand{\fov}{FoV\xspace}

\newcommand{\vhe}{VHE\xspace}

\newcommand{\gw}{\ensuremath{\txt{GW}}\xspace}
\newcommand{\gws}{\ensuremath{\txt{GWs}}\xspace}

\newcommand{\grb}{\ensuremath{\txt{GRB}}\xspace}
\newcommand{\grbs}{\ensuremath{\txt{GRBs}}\xspace}
\newcommand{\llgrb}{\ensuremath{\txt{LL-GRB}}\xspace}
\newcommand{\llgrbs}{\ensuremath{\txt{LL-GRBs}}\xspace}

\newcommand{\hlgrbs}{\ensuremath{\txt{high-luminosity GRBs}}\xspace}

\newcommand{\ebl}{\ensuremath{\txt{EBL}}\xspace}
\newcommand{\liso}{\ensuremath{{L_{\gamma,\text{iso}}}}\xspace}

\newcommand{\gamray}{\ensuremath{\gamma\txt{-ray}}\xspace}
\newcommand{\gamrays}{\ensuremath{\gamma\txt{-rays}}\xspace}

\newcommand{\irfs}{IRFs\xspace}

\renewcommand{\pl}{\ensuremath{\txt{PL}}\xspace}
\newcommand{\pls}{PLs\xspace}
\newcommand{\ecut}{\ensuremath{E_{\txt{cut}}}\xspace}

\newcommand{\lstm}{LSTM\xspace}
\newcommand{\lstms}{LSTMs\xspace}
\newcommand{\rnn}{RNN\xspace}
\newcommand{\rnns}{RNNs\xspace}

\newcommand{\andet}{anomaly detection\xspace}
\newcommand{\Andet}{Anomaly detection\xspace}
\newcommand{\cls}{classification\xspace}
\newcommand{\Cls}{Classification\xspace}

\newcommand{\pPoisAdB}{\ensuremath{p^{\mrm{bck}}_{\mrm{anm}}}\xspace}
\newcommand{\pPoisAdS}{\ensuremath{p^{\mrm{sig}}_{\mrm{anm}}}\xspace}
\newcommand{\pPoisAdBs}{\ensuremath{p^{\mrm{bck/sig}}_{\mrm{anm}}}\xspace}

\newcommand{\lambPois}{\ensuremath{\lambda}\xspace}
\newcommand{\lambPredB}{\ensuremath{\bm{\lambda}^{\mrm{bck}}_{\epsilon}}\xspace}
\newcommand{\lambPredS}{\ensuremath{\bm{\lambda}^{\mrm{sig}}_{\epsilon}}\xspace}
\newcommand{\lowSigRate}{\ensuremath{\bm{\kappa}_{\epsilon}}\xspace}

\newcommand{\ts}{\ensuremath{\mrm{TS}}\xspace}
\newcommand{\tsAd}{\ensuremath{\mrm{TS}_{\mrm{anm}}}\xspace}
\newcommand{\tsCls}{\ensuremath{\mrm{TS}_{\mrm{clas}}}\xspace}
\newcommand{\tsCtl}{\ensuremath{\mrm{TS}_{\mrm{ctl}}}\xspace}

\newcommand{\featBt}{\ensuremath{\bm{\phi}^{\mrm{enc}}_{\epsilon,\tau}}\xspace}
\newcommand{\featSt}{\ensuremath{\bm{\phi}^{\mrm{dec}}_{\epsilon,\tau}}\xspace}
\newcommand{\featBsT}{\ensuremath{\bm{\phi}_{\epsilon,\tau}}\xspace}

\newcommand{\featS}{\ensuremath{\bm{\phi}^{\mrm{dec}}_{\epsilon}}\xspace}
\newcommand{\predS}{\ensuremath{\bm{\psi}_{\epsilon}}\xspace}
\newcommand{\predClst}{\ensuremath{\bm{\zeta}_{\tau}}\xspace}
\newcommand{\predCls}{\ensuremath{\zeta}\xspace}

\newcommand{\predClsB}{\ensuremath{\zeta_{\mrm{bck}}}\xspace}
\newcommand{\predClsS}{\ensuremath{\zeta_{\mrm{sig}}}\xspace}

\newcommand{\kde}{KDE\xspace}
\newcommand{\kdes}{KDEs\xspace}
\newcommand{\bandW}{\ensuremath{h_{\mrm{KDE}}}\xspace}

\newcommand{\pdet}{\ensuremath{p_{\mrm{det}}}\xspace}
\newcommand{\fdet}{\ensuremath{f_{\mrm{det}}}\xspace}

\begin{document}
  \title{\Large{
    Deep learning detection of transients
  }}
  \shorttitle{
    Deep learning detection of transients
    \ifthenelse{\boolean{isDraft}}{
      \textcolor{hotpink}{\bf{- {\sffamily DRAFT} - \today~(\currenttime) -}}
    }{}
  }

  \author{Iftach Sadeh}
  \email{iftach.sadeh@desy.de}
  \altaffiliation{Deutsches Elektronen-Synchrotron (DESY),\\ Platanenallee 6, 15738 Zeuthen, Germany.}
  \nocollaboration

  \shortauthors{
    I.~Sadeh
    \ifthenelse{\boolean{isDraft}}{
      \textcolor{hotpink}{\bf{- {\sffamily DRAFT} - \today~(\currenttime) -}}
    }{}
  }

  \date{February 2019}

  \begin{abstract}
    The next generation of observatories will facilitate
    the discovery of new types of
    astrophysical transients.
    The detection of such phenomena,
    whose characteristics are presently poorly constrained,
    will hinge on
    the ability to perform blind searches.
    We present a new algorithm for this
    purpose, based on deep learning.
    We incorporate two
    approaches, utilising \andet and \cls techniques.
    The first is model-independent,
    avoiding the use of background modelling and
    instrument simulations.
    The second method enables
    targeted searches, relying on generic spectral
    and temporal patterns as input.
    We compare our methodology with the
    existing approach to serendipitous
    detection of gamma-ray transients.
    The algorithm
    is shown to be more robust, especially for
    non-trivial spectral features.
    We use our framework to derive the
    detection prospects of
    low-luminosity gamma-ray bursts with the
    upcoming Cherenkov Telescope Array.
    Our method is an unbiased, completely data-driven
    approach for multiwavelength and multi-messenger
    transient detection.
  \end{abstract}

  \ifthenelse{\boolean{isDraft}}{
    \textcolor{hotpink}{\bf{- {\sffamily DRAFT} - \today~(\currenttime) -}}
  }{}

  \section{Introduction}\label{SECintroduction}
  %
    Transient astrophysical phenomena at
    high energies present us with the opportunity
    to explore a broad range of fundamental physics.
    Recent years have seen a revolution of the field,
    with the first joint
    detections of gamma rays (\gamrays) and
    gravitational waves (\gws) \citep{PhysRevLett.119.161101};
    evidence for \gamray association with
    neutrinos \citep{IceCube:2018cha};
    and the first detection of a \gamray burst (\grb)
    at sub-\tev energies \citep{2019ATel12390....1M}.
    This has led to a wealth of scientific output,
    touching \eg on
    the formation of heavy elements;
    the physics of \grbs;
    and the properties of
    accelerators of cosmic rays \citep{2017ApJ...848L..13A, 2018A&A...615A.132R, 2018arXiv180601624R}.

    One of the interesting source populations that might
    be fully unveiled in the near future is that of
    low-luminosity \gamray bursts
    (\llgrbs) \citep{Liang:2006ci, Virgili:2008gp}.
    \llgrbs are a sub-class of the population
    of long \grbs, which has been connected
    to mildly relativistic supernovae \citep{Cano:2016ccp}.
    Compared to most bursts,
    \llgrbs are distinguished by
    low isotropic equivalent luminosities,
    generally considered for
    ${\powB{46} < \liso < \powB{49}~\erg\,\sec^{-1}}$.
    Only a small number of \llgrbs have been observed to date.
    However, there are indications that their observable rate in the
    local Universe (redshift, ${z < 0.1}$)
    is high, of the order of
    ${150\,\mrm{Gpc}^{-3}\,\mrm{yr}^{-1}}$ \citep{Sun:2015bda}.

    The exact nature of the
    prospective very-high-energy (\vhe)
    emission from \llgrbs is uncertain.
    However, the motivation to search for such events is high.
    For one, \llgrbs may be detectable as sources of
    gravitational waves \citep{Howell:2010hm}.
    Additionally, compared to their higher-luminosity counterparts,
    \llgrbs are favoured as the sources of ultra high energy cosmic rays.
    Primarily, this is because \llgrbs present an optimal balance between
    processes of particle acceleration and of energy
    loss \citep{2008PhRvD..78b3005M, 2018PhRvD..97h3010Z, 2018arXiv180807481B}.
    They therefore provide
    an efficient environment for cosmic ray generation and escape.
    Correspondingly,
    \llgrbs represent a unique multi-messenger window into
    the extreme energy regimes of \gamrays, cosmic rays, neutrinos, and \gws.

    \newpage
    Existing ground- and space-based
    observatories\footnote{\Eg~\fermi:~\url{https://fermi.gsfc.nasa.gov/}; \hess:~\url{https://www.mpi-hd.mpg.de/hfm/HESS/HESS.shtml}; \magic:~\url{https://wwwmagic.mpp.mpg.de/}; \veritas:~\url{https://veritas.sao.arizona.edu/}.}
    are capable of detecting \gamrays 
    within the regime relevant for \llgrbs.
    However, their current
    detection is very challenging.
    For instance,
    \fermig is not optimised for the detection of
    \llgrbs, due to their low peak synchrotron
    energies \citep{Sun:2015bda};
    for \fermil, detection is unlikely given
    the poor sensitivity
    to transients on short time scales \citep{Abdollahi:2016rso}.
    Conversely,
    imaging atmospheric Cherenkov telescopes (\iacts)
    can perform well on short intervals.
    However, the sensitivity of the current generation
    of instruments
    to transient phenomena below~${100~\gev}$
    is very limited.

    The upcoming Cherenkov Telescope Array
    (\cta)\footnote{\cta:~\url{https://www.cta-observatory.org/}.}
    will significantly improve upon the current facilities.
    Specifically, \cta will provide
    a large field-of-view (\fov) and improved
    energy resolution. It will also
    allow observation of \gamrays
    down to~${\Sim20~\gev}$, which will be critical
    for \llgrbs.
    The rate of observable \llgrbs under
    the duty-cycle of \cta is estimated
    as up to $1$~detection per-month
    (based on their relative
    number density with regards to \hlgrbs \citep{Inoue:2013vy}).
    As such, they are appealing
    targets for real-time searches.
     
    \llgrbs and 
    other putative transient populations
    are not well
    constrained by current observations.
    In order to maximise the potential
    of online searches, the
    corresponding algorithms will need to make
    as few model assumptions as possible.
    Accordingly,
    data-driven analysis methods have the potential
    to significantly enhance the existing infrastructure.

    In the following study we present a new method,
    intended for the detection of transient events.
    Our algorithm is based on deep learning,
    using recurrent neural networks. It employs
    two complementary approaches, optimised for both
    model-independent and targeted searches.

    Both training and evaluation of our
    estimator are computationally inexpensive,
    allowing for detection of transients on second
    time scales with insignificant latency.
    Our method can therefore be incorporated as part of
    real-time analyses.
    As such, it will be used by \cta to transmit
    transient event alerts to the astronomical community.
    These alerts will
    facilitate effective multiwavelength /
    multi-messenger followup of the discovered events,
    which will be crucial for their interpretation.

    We use our algorithm to make the first predictions
    for serendipitous \vhe detection of \llgrbs.
    We perform a scan
    of the parameter space of events,
    identifying those that would be within reach of \cta.
    Compared to the existing state of the art,
    we achieve higher detection rates
    for non-trivial source types.

    While we illustrate our new approach
    using \gamray sources, it is by design generic and
    model independent.
    The methodology is not restricted to a specific energy regime,
    type of input, or time scale.
    In particular, it is well suited for multiwavelength
    and multi-messenger searches,
    where different observable are combined.
    It can therefore easily be adapted
    for many other transient searches.

  \section{Existing detection methodologies}\label{xxx}
    %
    \subsection{Techniques for \gamray source detection}\label{SECexistingTechniquesForDetection}
      %
      \iacts must exclude a
      significant amount of background, mostly
      originating from cosmic rays \citep{Berge2007, funk_review}.
      Unless extremely short time intervals are considered,
      this background is irreducible, and must
      be corrected for on average.
      %

      Different approaches are used in order
      to identify excesses of reconstructed events.
      One common method is to define \tit{on-}
      and \tit{off-regions} of observation within the
      same \fov \citep{2001A&A...370..112A, Berge2007}. For
      this kind of analysis, a source is searched for
      within the on-region, while the off-region(s) are
      assumed to only contain background events.
      Contemporaneous measurements
      within off-regions are used to
      estimate the number density of background events,
      which is then subtracted from counts in the on-region.

      This type of method has the advantage that the background
      is derived directly from the data. 
      However, the off-regions are by construction separated from
      the position of the source. The method
      is therefore susceptible to
      uncertainties on \eg the homogeneity of the
      performance of the camera.

      Another approach is to perform a
      likelihood analysis, which requires explicit modelling
      of both the source and the background \citep{Knodlseder:2016nnv}.
      In a common application of this
      technique, the background is estimated by a fit
      over the entire \fov, covering an
      area where no sources are expected to exist.

      Such methodology is very powerful, as it naturally
      allows for sophisticated background modelling.
      However, it has the disadvantage of strongly depending on
      the accuracy of
      instrument response functions (\irfs), 
      which encapsulate the
      effective area, radial acceptance,
      point spread function 
      and energy dispersion of the instrument.

      For both approaches,
      one must usually make
      additional assumptions on
      the spectral and temporal properties of
      the source \citep{2017APh....93....1W}. This may 
      limit the discovery potential for unexpected transient
      types.

    \subsection{Transient detection strategies}
    %
      The nominal strategy for observing \grbs with \cta
      is to follow external alerts, which would be generated by
      other instruments \citep{Acharya:2017ttl}.
      This is mainly driven by the low rate
      of occurrence of \hlgrbs \citep{Inoue:2013vy},
      which makes their serendipitous discovery by \cta unlikely.

      In the case of \llgrbs, the potentially high rates motivate
      blind searches. These would be performed within the \fov
      of \cta during normal operations,
      not as a response to an alert.
      We follow this approach exclusively in the current study.
      Our nominal strategy is to conduct
      continuous searches.
      In the most simple scenario, these
      would be performed in multiple independent
      regions of interest (\rois), covering the
      entire \fov.

      Real-time discovery of transients with \iacts
      is challenging.
      Existing methods have to control a
      large number of effects, such as
      undetected sources within the \fov;
      imperfect modelling of galactic foregrounds
      or of the effect of stars;
      and uncertainties on the
      combination of data from multiple
      observation epochs (which may be necessary
      for background estimation).
      The latter in particular 
      implies that one must account for
      many parameters, \eg
      the zenith and azimuth angles
      of observation; the night sky background;
      and
      changes in the density and transparency
      of the atmosphere \citep{performance_stereo_MAGIC}.

      It is foreseen that \cta will use dedicated \irfs
      for min--hour observation
      periods, which would account for
      changing observing conditions.
      Such simulations
      would be generated with some delay after data taking.
      Correspondingly, the online analysis will
      be based on \irfs compatible with
      averaged anticipated conditions.

      The focus of this paper
      is the development of new detection methods. In light of
      the challenges discussed above, our objective
      is to minimise the dependence on
      modelling, as well as the influence of observational effects.

  \section{New detection methods}
    %
    This study presents two novel transients detection algorithms,
    denoted in the following as
    \tit{anomaly detection}, and
    \tit{classification}.

    \Andet represents a model-independent
    approach, where transient events are identified based on
    their divergence from the expected background.
    This simple methodology is completely data-driven, and
    is able to adapt to real-time evolution of the background.

    Our approach has a clear
    advantage over traditional methods;
    the background model is derived \insitu, and
    from the same \roi as for the source.
    \Andet is therefore completely decoupled from the simulation
    of the instrument.
    It is also insensitive to uncertainties on
    modelling of the background,
    the atmosphere,
    or any of the other artefacts mentioned above.

    Our second new method employs classification.
    In this case, we
    train an estimator to identify transient patterns.
    While less model-independent than \andet, the
    \cls approach increases the sensitivity of specific searches,
    such as for \llgrbs.

    Compared to existing methods, \cls has several advantages.
    Similarly to \andet, data from different \rois
    need not be mixed. In addition,
    the time structure of transient events is
    naturally incorporated as a part
    of the training process,
    avoiding the need for explicit modelling.
    Finally, simple training examples may be used
    as the basis for detecting sources with more complicated intrinsic
    spectra, as illustrated below.

    \subsection{Recurrent neural network estimator}
      %
      \begin{figure*}[tp]
        \begin{minipage}[c]{1\textwidth}

          \begin{minipage}[c]{1\textwidth}
            \begin{center}
              \colorbox{figColBox}{\includegraphics[trim=16mm 90mm 49mm 10mm,clip,width=1\textwidth]{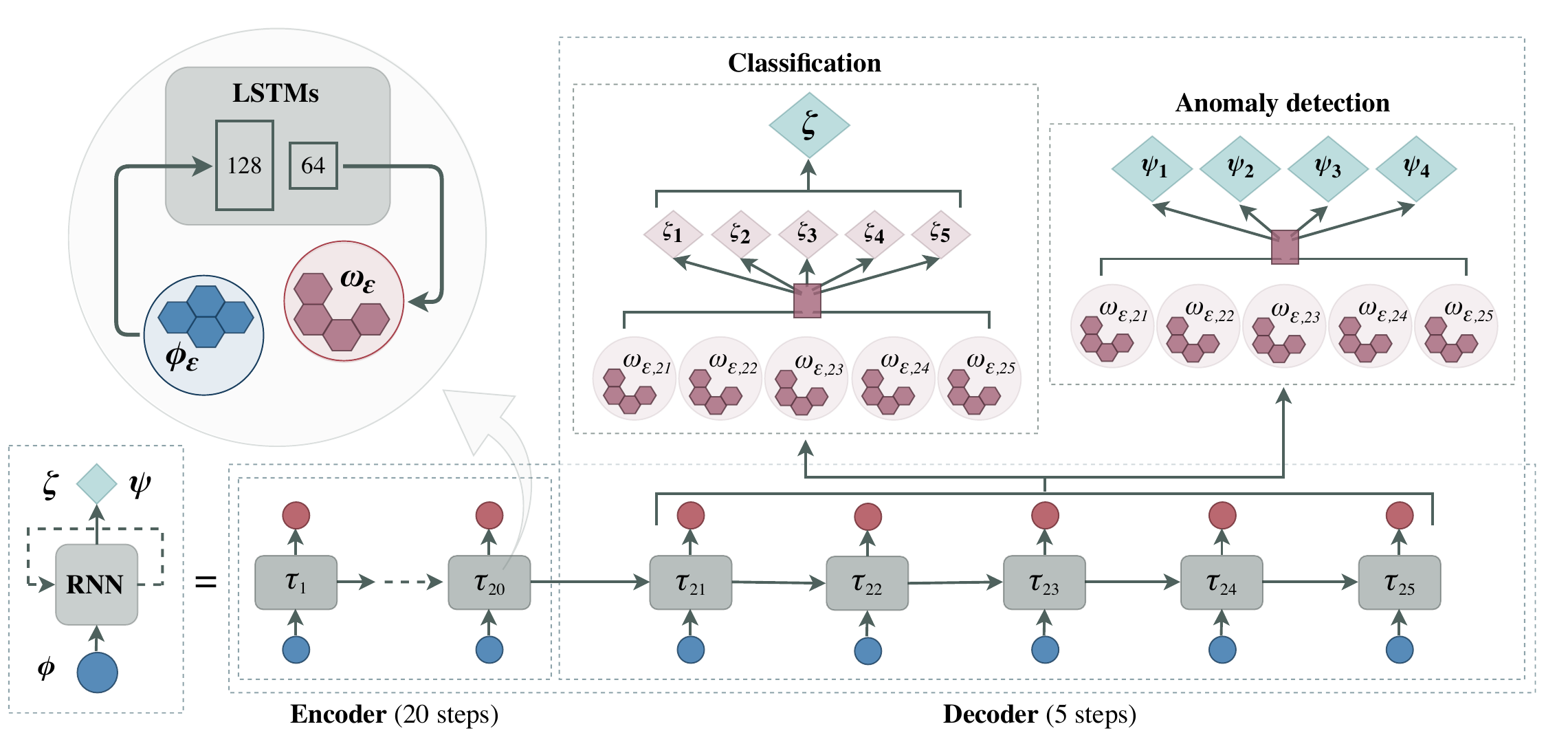}}
            \end{center}
          \end{minipage}\hfill
          %
          %
          \begin{minipage}[c]{1\textwidth}
            \begin{center}
              \begin{minipage}[t]{1\textwidth}\begin{center}
                \caption{\label{FIGannArch}Schematic
                  design of the \rnn used in this study.
                  The unrolled representation may be decomposed into
                  an encoder and a decoder. These respectively
                  represent~$20$ and~$5$ time steps, $\tau$,
                  of \lstm units (rectangles).
                  Each \lstm comprises two layers
                  of $128$ and $64$ hidden units.
                  The input data, $\phi$, (blue circles)
                  make up $4$ numbers for each time
                  step (blue hexagons), corresponding to event counts in
                  four energy bins, $\epsilon$.
                  The direct output of the \lstms,
                  $\omega_{\epsilon,\tau}$, (red hexagons) are predictions
                  for event counts for each step and energy bin.
                  The output of the \rnn (diamonds)
                  depends on the implementation.
                  For the case of \andet,
                  the output, $\psi_{1-4}$, provides the predicted
                  background counts for each of the energy bins,
                  integrated over the decoder time steps. The output of the
                  \cls method, $\zeta$,
                  may be calibrated into a probability density function.}
              \end{center}\end{minipage}\hfill
            \end{center}
          \end{minipage}\hfill
          \vspace{15pt}
        \end{minipage}\hfill
      \end{figure*} 
      %

      Machine learning is widely used in astronomy
      \citep{Sadeh:2015lsa, 2017APh....89....1K,2018MNRAS.476.2117R}.
      Deep learning, and in particular convolutional neural networks,
      have shown great promise.
      Successful applications include
      optical and \gamray object classification,
      and transient searches in
      \eg images and radio signals
      \citep{2018APS..APRL01031K, 2019MNRAS.484...93D, 2018MNRAS.476.5365S, 2017MNRAS.472.3101G, Erdmann:2019nie}.

      In the current study, we utilise
      a recurrent neural network (\rnn),
      made up of long short-term memory (\lstm) units.      
      \rnns are a type of artificial neural network,
      which is well suited for time series analysis.
      They have numerous applications, ranging from
      natural language translation
      \citep{Sutskever:2014:SSL:2969033.2969173}
      to denoising of \gw signals
      \citep{Shen:2017jkj}.

      The connections between the \lstm nodes 
      in an \rnn form
      a directed graph, representing a sequence of steps in time.
      Outputs from each time step are fed as input to the next,
      in addition to the respective temporal data.

      In principle, \rnns may be used to make predictions
      for arbitrarily distant inputs.
      For computational reasons, it is common to implement
      \tit{unrolled} versions of \rnns. These contain
      a fixed number of steps, \ie a fixed number
      of inputs and outputs.
      For a review of deep learning and \rnns, see~\cite{deepLearningReview}.

      The neural network used in this study is
      implemented with the open-source
      software, \tensorflow \citep{tensorflow2015-whitepaper}.
      The architecture of the \rnn is
      illustrated in \autoref{FIGannArch}.

      The network accepts an input which
      corresponds to $25$~time steps, each representing
      a~${1~\sec}$ interval of \gamray data.
      %
      The different steps are implemented as \rnn cells. A cell
      is composed of a pair of \lstm layers,
      respectively comprising~$128$ and~$64$
      hidden units. The hidden units
      are conceptually similar to nodes in a feedforward neural network.
      As such, they represent the set of parameters
      which are tuned during training.

      The network may be decomposed into two elements, an \tit{encoder} and a
      \tit{decoder}. The encoder receives $20$~time steps
      as input.
      A potential transient signal event is then
      searched for within the $5$~time steps associated with the
      decoder.

      The input data corresponding to a given time step 
      are a list of \tit{features}. These are
      respectively denoted by
      ${\{\featBt\}_{\tau \in 1-20}}$ and
      ${\{\featSt\}_{\tau \in 21-25}}$,
      for the encoder and decoder time steps.
      In the current
      study, the features are event counts in $4$
      logarithmically-spaced energy bins within
      ${30 < E_{\mrm{\gamma}} < 200~\gev}$
      (starting slightly above the
      lower energy threshold of \cta).

      The inputs to the encoder
      are assumed to correspond to background-only
      counts in all cases.
      The input to the decoder and the
      output of the network depend on the
      type of inference being used, as discussed below.

      We did not explore the full parameter space of
      possible \rnn architectures.
      Rather, the chosen temporal and energetic representation
      is motivated by the expected properties of \llgrbs, and
      is intended to illustrate our methodology.
      %
      It is possible that better performance could be
      achieved with a different configuration, which we leave
      for future work.

      We also note that the architecture is easily generalisable
      to different time scales. It may also
      incorporate additional
      features, such as data-quality metrics; 
      per-event background rejection probabilities;
      information related
      to the weather; optical observations of the perspective
      transient; and detected artefacts (\eg meteors).
      Such observables may take the form of 
      single numbers, probability density functions,
      CCD images, \etc

    \subsection{\Andet}
      %
      Using \andet, 
      transient events are identified by detecting significant
      deviations of observed event counts, compared
      to the expected background.
      Poissonian statistics are assumed
      for both the background and the signal models.\footnote{
      For brevity, we refer to the
      background-only hypothesis as the \tit{background model},
      and to the background$+$signal hypothesis
      as the \tit{signal model}.}
      The corresponding probability distributions are
      \pPoisAdB and \pPoisAdS, where
      ${\pPoisAdBs(k \vert \lambPois) = e^{-\lambPois}} \lambPois^{k} / k!$,
      given the \tit{rate parameter},~\lambPois.

      The background in this case is derived \insitu. This is done
      using data exclusively from within the
      \roi for the source.
      In the most simple application,
      spectral modelling of the source is not needed,
      and a simple top-hat function is assumed
      for the temporal behaviour.
      More sophisticated assumptions on the source
      may be taken. For instance, one could
      weight the event counts of the different time steps by an
      assumed temporal trend \citep{2017APh....93....1W}.
      We do not take such an approach here, as
      this would
      potentially limit the detectability
      of unexpected transient patterns.

      The network is trained using background-only 
      events for all $25$~time steps.
      Training events can in general be derived from simulations,
      or from historical or near-contemporaneous data.
      This methodology is particularly powerful,
      as the estimator will continuously be 
      optimised (retrained) using
      real-time data.

      The network is trained by optimizing
      (minimising) a \tit{loss function}. The latter encodes
      the difference between the output of the network
      and the intended outcome.
      The purpose of the \andet \rnn is to
      directly predict the number of
      counts inside the \roi.
      In the current study, we chose to integrate the
      output of the decoder
      across the $5$~time steps.
      The corresponding predictions for the~$4$
      energy bins are denoted by ${\{\predS\}_{\epsilon \in 1-4}}$.
      The loss function for training is therefore defined as
      the absolute difference between 
      observed event counts and \predS predictions.

      When the trained network is used to evaluate
      data, it receives a sequence of~$25$ input features, \featBsT.
      In the case that a transient event occurs, the
      signal features, \featSt, (time steps, ${\tau_{21} - \tau_{25}}$)
      may correspond to higher event rates.
      Such a pattern is never introduced as part of the
      training, and may
      be incompatible with the background-only model of the \rnn.
      We therefore replace the~$5$ steps, \featSt,
      with inputs from the encoder.
      The predictions from the trained network are then
      used to estimate the rate parameters of the
      Poissonian background distributions,
      ${\{\lambPredB\}_{\epsilon \in 1-4}}$.

      Up to this point in the analysis,
      we have only utilised data from the encoding
      phase of the network (${\tau_{1} - \tau_{20}}$).
      In order to estimate the Poissonian probability function of the
      signal model, we use
      \featS. The latter stand for the
      counts of the perspective signal in each energy bin,
      integrated over time steps, ${\tau_{21} - \tau_{25}}$.

      The Poissonian rate parameters of the signal
      are estimated for the different energy bins as
      \begin{equation}
        \lambPredS = \max\{\featS,\, \lowSigRate\}.
      \label{eqLambPredS} \end{equation}
      \noindent
      The parameter, \lowSigRate, is nominally selected as
      ${\lowSigRate = \lambPredB}$ in the current study.
      Setting the background rate as a lower limit for the signal
      ensures that downward fluctuations of \featS
      moderate the significance of a detection; this
      reduces the rate of spurious detections by a large fraction.
      It is also possible to use higher values
      of \lowSigRate, corresponding to increasingly conservative
      detection thresholds. For instance, one may choose to set
      ${\lowSigRate = (\lambPredB + n_{\mrm{min}})}$.
      On average, this would be akin to
      requiring that at least ${n_{\mrm{min}}}$ \gamrays above background are
      detected in each energy bin.

      The test statistic for detecting a transient
      signal within the ${\tau_{21} - \tau_{25}}$ time interval
      can finally be derived, reading
      \begin{equation}
        \tsAd = 
        -2 \log
        \left( 
          \frac {\pPoisAdB(\featS \vert \lambPredB)}
                {\pPoisAdS(\featS \vert \lambPredS)}
        \right).
      \label{eqTsAd} \end{equation}

    \subsection{\Cls}
      %
      Instead of using the \rnn to predict background
      counts within different temporal/energy bins,
      the estimator may be used to directly classify transient
      events. In this case, an external layer
      is added to the decoder, mapping the output
      (of event counts)
      into $5$ \tit{logits}, denoted by \predClst.
      Each logit represents a probability density function
      for an event to belong
      to the signal class,
      based on a single decoder time step
      (within ${t_{21} - t_{25}}$).

      The network is trained using labelled examples
      of background and signal events. Correspondingly,
      the output logits represent the inferred probability
      that an event belongs to the signal class.
      The loss function which is optimised during
      training represents the
      accuracy of all logits. The combined
      accuracy is calculated as a weighted average across
      the~$5$ time steps,
      where the weights are themselves optimised during training.
      Consequently, the unique time structure of signal events
      drives the training.

      The output of the \rnn, \predCls, is the inferred
      classification metric for a given event
      (see \autoref{FIGclsDist} below).
      We define the test statistic for identifying
      a signal event 
      as
      \begin{equation}
        \tsCls = -2 \log \left(
          \frac{\predClsB}{\predClsS}
        \right),
      \label{eqTsCls} \end{equation}
      \noindent
      following the prescription of~\cite{Cranmer:2015bka}.
      Strictly speaking, the combined
      classification metric, \predCls, is not a logit
      (for computational reasons it is
      not normalised to span the interval~${[0,1]}$).
      However, the distributions of \predCls
      can be calibrated to serve as
      probability density functions for each class.

      In order to prevent distortions due to binning effects,
      we do not use the value of \predCls directly.
      Rather, we first fit the distributions of \predCls
      with kernel density estimators (\kdes).
      A \kde is a non-parametric method for parametrising
      probability density functions \citep{parzen1962}.
      It depends only on a
      single \tit{bandwidth} parameter, \bandW.
      The bandwidth defines a smoothing
      scale for the distributions of \predCls,
      and is optimised to the results of the training.

  \section{\grb simulations}
    %
    We demonstrate the application of our algorithm, by
    making predictions
    for serendipitous detection of \llgrbs with \cta,
    as described in the following.

    \subsection{Spectral models}
      %
      Due to their low luminosities, \llgrbs are expected to
      be characterised by low bulk Lorentz factors (of the order of~$10$),
      and low values of their peak spectral energy
      distribution
      (${\Sim10 - 100~\kev}$ in the observer frame)
      \citep{Ghirlanda:2017opl}.
      Consequently, their synchrotron emission is not
      likely to be detectable at multi-\gev energies.
      While a higher-energy inverse-Compton component might
      in fact be observable by \cta, it is also
      possible that this signal is suppressed
      due to absorption inside the source
      \citep{RudolphLlgrbsInPrep}.
      Despite these uncertainties, it remains important
      to perform \vhe searches for \llgrbs.
      Any detection will significantly advance
      our understanding.

      In order to simulate the possible \gamray signatures of \llgrbs,
      we use the following reference events:
      GRBs~${\mrm{080916C}}$, ${\mrm{090323}}$,
       ${\mrm{090510}}$, ${\mrm{090902B}}$,
       and~${\mrm{110731A}}$. 
      These are all bright \hlgrbs, which have been detected at
      high energies with \fermil (see ~\cite{Ackermann:2013zfa}
      and references therein).

      The selected reference events are best fit
      by different types of models
      \eg a Band function \citep{1993ApJ...413..281B},
      a power law (\pl), or a \pl with
      an exponential cutoff.
      It is therefore possible that the spectral
      components detected by \fermil are in fact a part of the
      afterglow, rather than
      of the prompt phase of these \grbs.
      In the following,
      we assume that this is not the case. That is,
      we interpret the \gev
      emission as
      an extension of \eg the Band model to high energies.
      In such a case, it is possible to predict the corresponding
      prompt \gev component of \llgrbs, based on
      a simplistic scaling
      of the flux (\cf \cite{Inoue:2013vy}).

      We begin by randomly
      shifting the reference \grbs in redshift and luminosity
      to the expected ranges for \llgrbs.
      In order
      to simulate the signals at \gev energies,
      we nominally assume a simple spectral/temporal \pl model,
      \begin{equation}
      M_{\mrm{PL}}(E,t) =
        k_{0}
        \left(\frac{E}{E_{0}}\right)^{-\Gamma}
        t^{-\tau} .
      \label{eqGrbPlModel} \end{equation}
      %

      The \tit{prefactor} and \tit{pivot energy},
      $k_{0}$ and $E_{0}$, are derived directly from the
      flux of the \grb.
      The spectral index, $\Gamma$, and temporal
      decay index, $\tau$, are randomly selected
      for each event, uniformly distributed
      within
      $1.9 < \Gamma < 2.7$
      and
      $0.8 < \tau < 2$.
      These properties
      generally correspond to the expectations
      for the low-luminosity population. We only consider
      those bursts which exhibit durations of the order of
      tens of seconds, excluding the
      population of ultra-long \grbs \citep{Levan:2013gcz}.

      We also simulate bursts
      having an exponential cutoff. The corresponding
      spectral models are parametrised as
      \begin{equation}
        M_{\mrm{EC}}(E) =
          M_{\mrm{PL}} \cdot
          \exp
          \left(
            - \frac{E}{\ecut}
          \right),
      \label{eqGrbExCutoffModel} \end{equation}
      \noindent
      where \ecut is the \tit{cutoff energy}.

      The observed spectra of cosmological \vhe sources
      are distorted by
      interactions with low-energy photons from the
      extragalactic background light (\ebl).
      The bursts considered in this study
      have low redshift and \gamray energies (compared
      to most \hlgrbs). Consequently,
      the effect of the \ebl is not expected to be important.
      We compared our results using several \ebl
      models \citep{Franceschini:2008tp, doi:10.1111/j.1365-2966.2010.17631.x, doi:10.1111/j.1365-2966.2012.20841.x}.
      We found that the effect of using a particular model or none
      at all is indeed insignificant for our sample.

    \subsection{Event simulation}
      %
      We generate \iact
      events using the open-source software,
      \ctools \citep{Knodlseder:2016nnv}.
      The latter is one of the proposed analysis frameworks
      for \cta, which implements the likelihood method
      discussed in \autoref{SECexistingTechniquesForDetection}.
      The Northern array of \cta is simulated
      using the publicly available \irfs (version \prodIrf).
      In the current study, we exclusively use \irfs
      optimised for ${30~\min}$ observations at zenith
      angles of~${\Sim20\dgr}$.
      We leave the comprehensive characterisation of
      \llgrb detectability under different observing conditions
      for future studies.

      We simulate two samples, one composed entirely
      of background counts, and the other including \grbs.
      All bursts are modelled as \pl spectra (\autoref{eqGrbPlModel}),
      unless otherwise indicated (\cf \autoref{FIGpdetExCutoff}).
      The energy range of
      \gamrays is restricted to reconstructed values,
      ${30 < E_{\gamma} < 200~\gev}$, where
      most of the emission from \llgrbs is expected
      to be detected.

      The \roi for the simulation is chosen as a circular
      region with a radius of ${0.25\dgr}$, centred 
      at the position of the source. The latter
      is displaced by ${0.5\dgr}$ from the centre
      of the \fov.
      For each sequence, we derive
      \featBsT by counting the number of
      reconstructed \gamray-events within the \roi
      for each time step and energy bin.

      The counts, \featBsT,
      are susceptible to fluctuation due
      to imperfect \gamray reconstruction, as well as to
      uncertainties on the \irfs.
      In particular, energy dispersion
      below~${\Sim50~\gev}$ may result in
      migration between bins, and can
      change the energy threshold of the analysis.
      We studied these effects by allowing~$10\%$ variation
      on the \irfs. We found that the propagated uncertainties
      do not significantly affect our results.

      As part of a realistic online analysis, searches would
      be performed as sliding windows in time.
      For the purpose of the current study,
      we only conduct searches over the ${5~\sec}$ intervals
      that coincide with the beginning of bursts in the signal sample.
      This is done regardless of the value
      of the randomised temporal decay
      parameter, $\tau$.
      We accordingly simulate a ${25~\sec}$ interval for each burst.
      The first ${20~\sec}$ (before the \grb emission would
      begin) are fed into the encoder of the \rnn (\featBt), and the next
      ${5}$ into the decoder (\featSt).

    \subsection{Detection significance}
      %
      Performing continuous blind searches for transient
      events incurs a
      penalty on the test statistic. To correct
      for the number of trials,
      we assume the following connection between
      pre- and post-trials probabilities,
      ${p_{\mrm{post}} = 1 - \left( 1 - p_{\mrm{pre}} \right)^{n_{\mrm{trial}}}}$ \citep{BILLER1996285}.

      The number of trials, ${n_{\mrm{trial}}}$, represents
      both the frequency of searches and the number of \rois
      within the \fov of \cta.
      The former corresponds
      to $100$~hours of observations
      at ${1~\sec}$ search intervals in the current study.
      The latter accounts for a conservative~${100}$
      simultaneous observations every second. In total,
      we consider \powA{3.6}{7} trials.

      We compare the results of our new
      algorithm with the
      test statistic computed by \ctools, \tsCtl,
      taken as a proxy for the likelihood methodology.
      The \tsCtl metric is derived exclusively
      from the final ${5~\sec}$ interval, during which the \grb
      would be active.

  \section{Results}
  %
    As the benchmark performance metric for simulated
    \grbs, we define the \tit{detectability},
    ${\pdet = \left< \rho_{\mrm{det}} \right>}$, for
    \begin{equation}
      \begin{matrix}
        %
        \rho_{\mrm{det}}(t) = 
        &
        \left\{ \begin{matrix} 
        0
        \;,
        &
        t < \ts_{5\sigma}
        \\
        1\;,
        &
        t \geq \ts_{5\sigma}
        \end{matrix} \right. .
        & \\
      \end{matrix}
    \label{eqGrbDetectability} \end{equation}
    \noindent
    Here $t$ represents the test
    statistic derived for a given detection method;
    ${\ts_{5\sigma}}$ is the corresponding
    threshold for a $5\sigma$ detection,
    where \eg for a model with a single degree
    of freedom, ${\ts_{5\sigma} = 25}$ \citep{Wilks:1938dza}.

    The detectability metric is averaged over 
    samples with different combinations of parameters
    (\eg spectral indices and redshifts).
    The absolute value of
    \pdet depends on the sample composition, which
    in this case is not physically motivated
    by a \grb luminosity function.
    However, \pdet may be used 
    to identify the most
    promising regions of the parameter space.
    It can also serve for
    objective comparison between the different detection
    methods.

    \begin{figure}[t]
      \begin{minipage}[c]{0.47\textwidth}

        \begin{minipage}[c]{1\textwidth}
          \begin{center}
            \colorbox{figColBox}{\includegraphics[trim=0mm 12mm 0mm 13mm,clip,width=.98\textwidth]{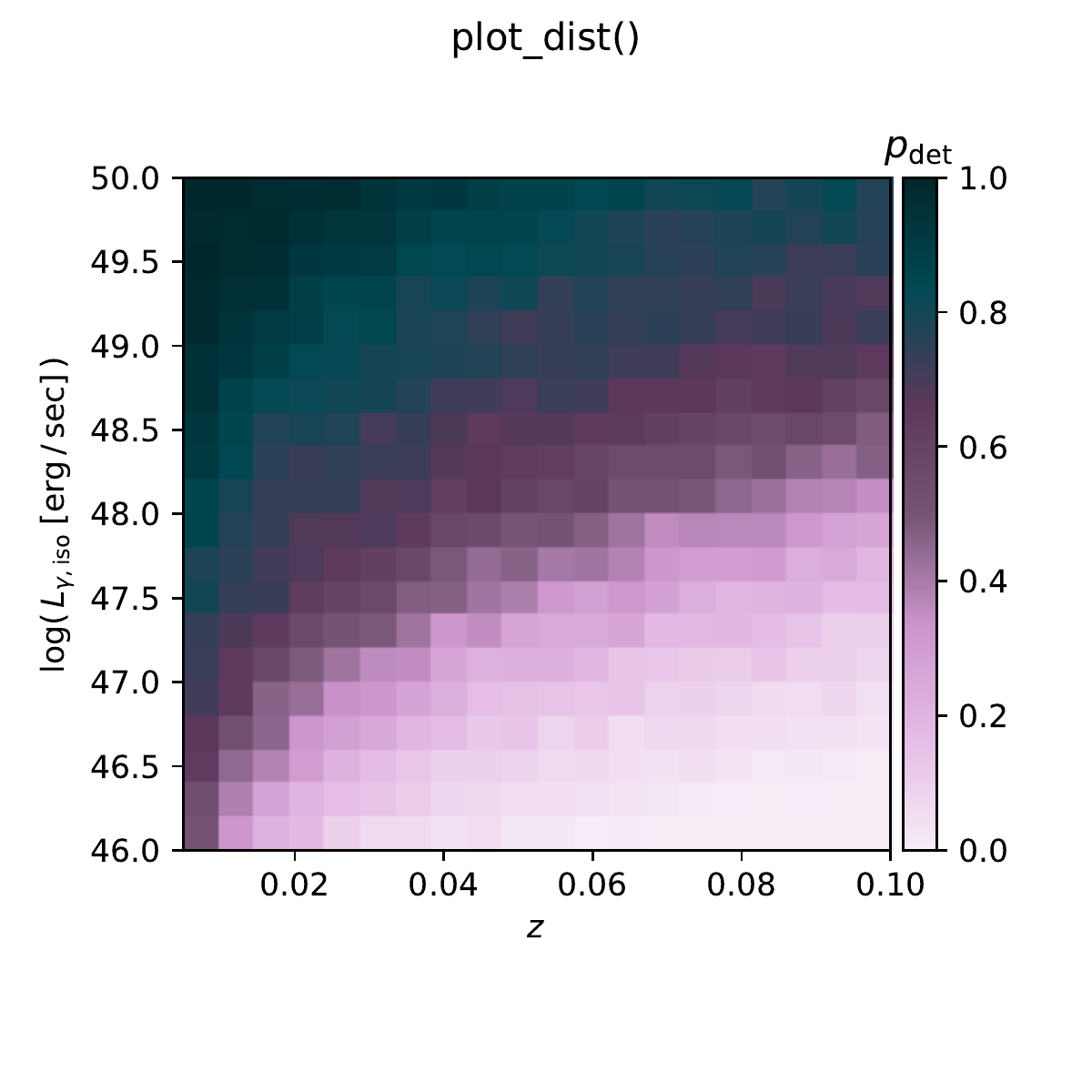}}
          \end{center}
        \end{minipage}\hfill
        %
        %
        \begin{minipage}[c]{1\textwidth}
          \begin{center}
            \begin{minipage}[t]{1\textwidth}\begin{center}
              \caption{\label{FIGpdetLisoVsZ}The
                detectability metric, \pdet, derived with
                \ctools, after accounting for trials.
                The simulation sample includes different
                combinations of redshift, $z$, and
                isotropic equivalent luminosity, \liso,
                spanning the expected properties of \llgrbs,
                assuming \pl spectral models.}
            \end{center}\end{minipage}\hfill
          \end{center}
        \end{minipage}\hfill
        \vspace{5pt}
      \end{minipage}\hfill
    \end{figure} 
    %
    The distribution of ${\pdet(\tsCtl)}$
    as a function of redshift, $z$, and
    isotropic equivalent luminosity, \liso,
    is shown in \autoref{FIGpdetLisoVsZ}.
    Much of the relevant parameter space
    of \llgrbs is available for \cta.
    Lower redshift values and higher
    \grb luminosities 
    are correlated with higher event fluxes.
    As expected, these also
    correspond to higher
    probabilities for a burst to be detected.

\begin{figure*}[tp]

      \begin{minipage}[c]{1\textwidth}

        \begin{minipage}[c]{1\textwidth}
          \begin{center}
            \begin{minipage}[c]{0.47\textwidth}\begin{center}
              \colorbox{figColBox}{\includegraphics[trim=0mm 12mm 0mm 18mm,clip,width=.98\textwidth]{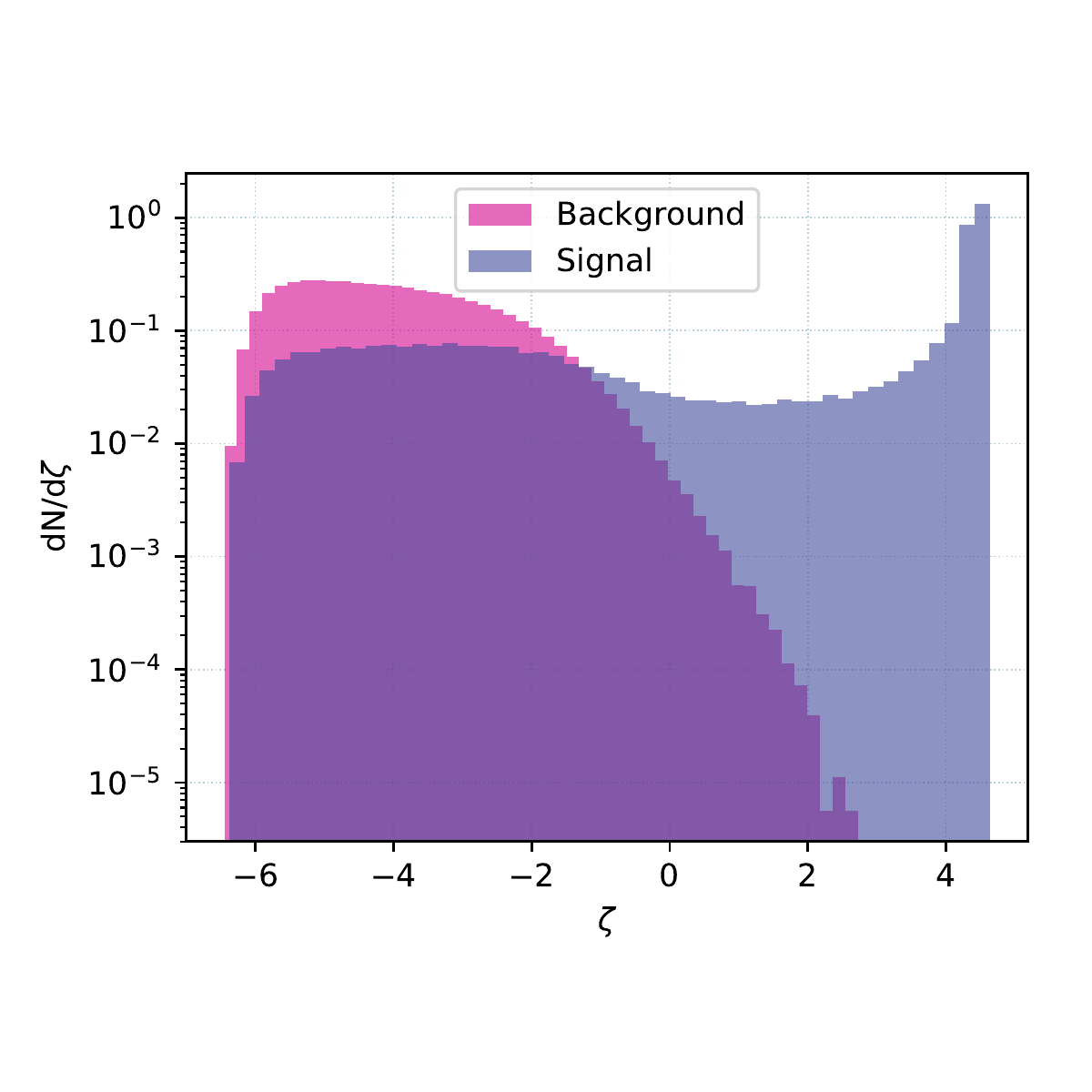}}
            \end{center}\end{minipage}\hfill
            \begin{minipage}[c]{0.47\textwidth}\begin{center}
              \colorbox{figColBox}{\includegraphics[trim=0mm 12mm 0mm 18mm,clip,width=.98\textwidth]{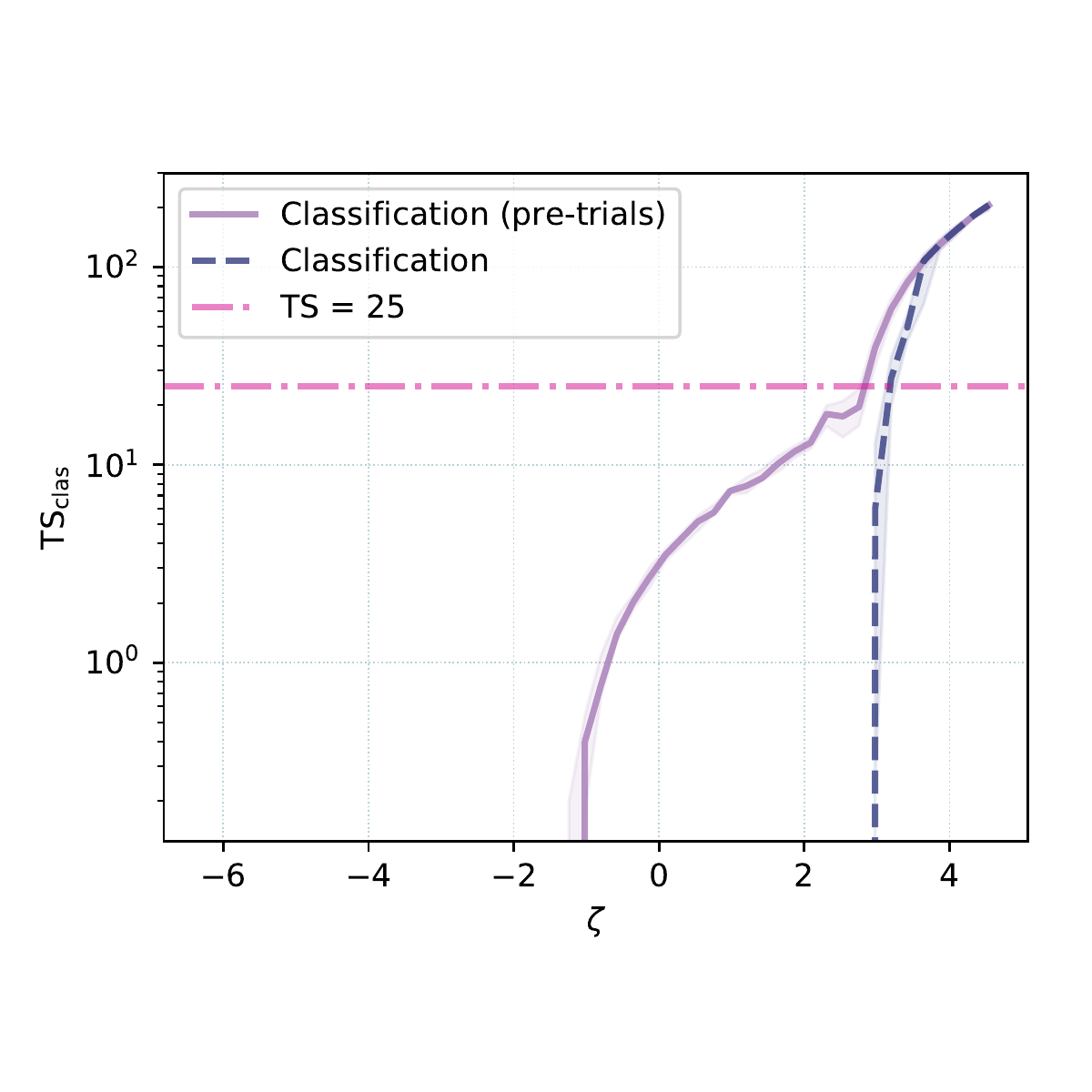}}
            \end{center}\end{minipage}\hfill 
          \end{center}
        \end{minipage}\hfill
        \vspace{5pt}
        \begin{minipage}[c]{1\textwidth}
          \begin{center}
            \begin{minipage}[c]{0.47\textwidth}\begin{center}
              \centering\subfloat[]{\label{FIGclsDist}}
            \end{center}\end{minipage}\hfill
            \begin{minipage}[c]{0.47\textwidth}\begin{center}
              \centering\subfloat[]{\label{FIGclsTs}}
            \end{center}\end{minipage}\hfill 
          \end{center}
        \end{minipage}\hfill
        %
        %
        \begin{minipage}[c]{1\textwidth}
          \begin{center}
            \begin{minipage}[t]{1\textwidth}\begin{center}
              \caption{\label{FIGclsPerformance}Parametrisation
                of the performance of the trained \cls method.
                \Subref{FIGclsDist}~Distributions of the
                \cls metric, \predCls, for the signal
                and background samples, as indicated.
                \Subref{FIGclsTs}~The parametrised
                \cls test statistic, \tsCls,
                (see \autoref{eqTsCls}) as a function of \predCls,
                before and after the correction for trials.
                The dashed-dotted horizontal line highlights the
                value, $\ts = 25$.}
            \end{center}\end{minipage}\hfill
          \end{center}
        \end{minipage}\hfill
        \vspace{15pt}
      \end{minipage}\hfill

    \end{figure*}

    Using the trained \rnn in the \cls mode,
    we derive the distributions of the
    \cls metric, \predCls, for the
    background and signal samples,
    as shown in \autoref{FIGclsDist}.
    The distributions are used to fit a \kde estimator.
    %
    We chose the \kde bandwidth such 
    that the resulting distribution
    of \tsCls is smooth for values~${> 10}$.
    The corresponding relation between \tsCls and \predCls
    is presented in \autoref{FIGclsTs}.

\begin{figure*}[tp]

      \begin{minipage}[c]{1\textwidth}

        \begin{minipage}[c]{1\textwidth}
          \begin{center}
            \begin{minipage}[c]{0.47\textwidth}\begin{center}
              \colorbox{figColBox}{\includegraphics[trim=0mm 12mm 0mm 18mm,clip,width=.98\textwidth]{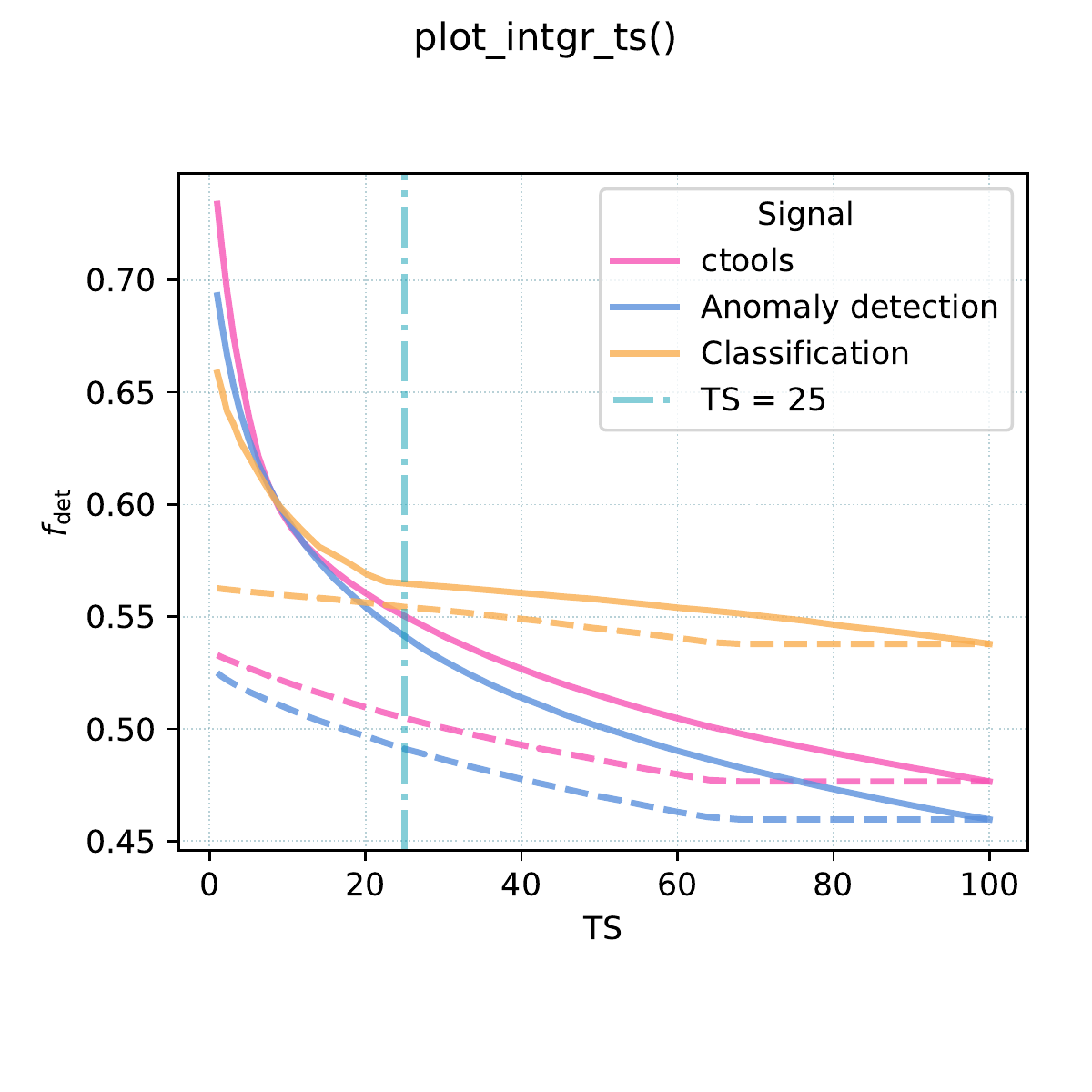}}
            \end{center}\end{minipage}\hfill
            \begin{minipage}[c]{0.47\textwidth}\begin{center}
              \colorbox{figColBox}{\includegraphics[trim=0mm 12mm 0mm 18mm,clip,width=.98\textwidth]{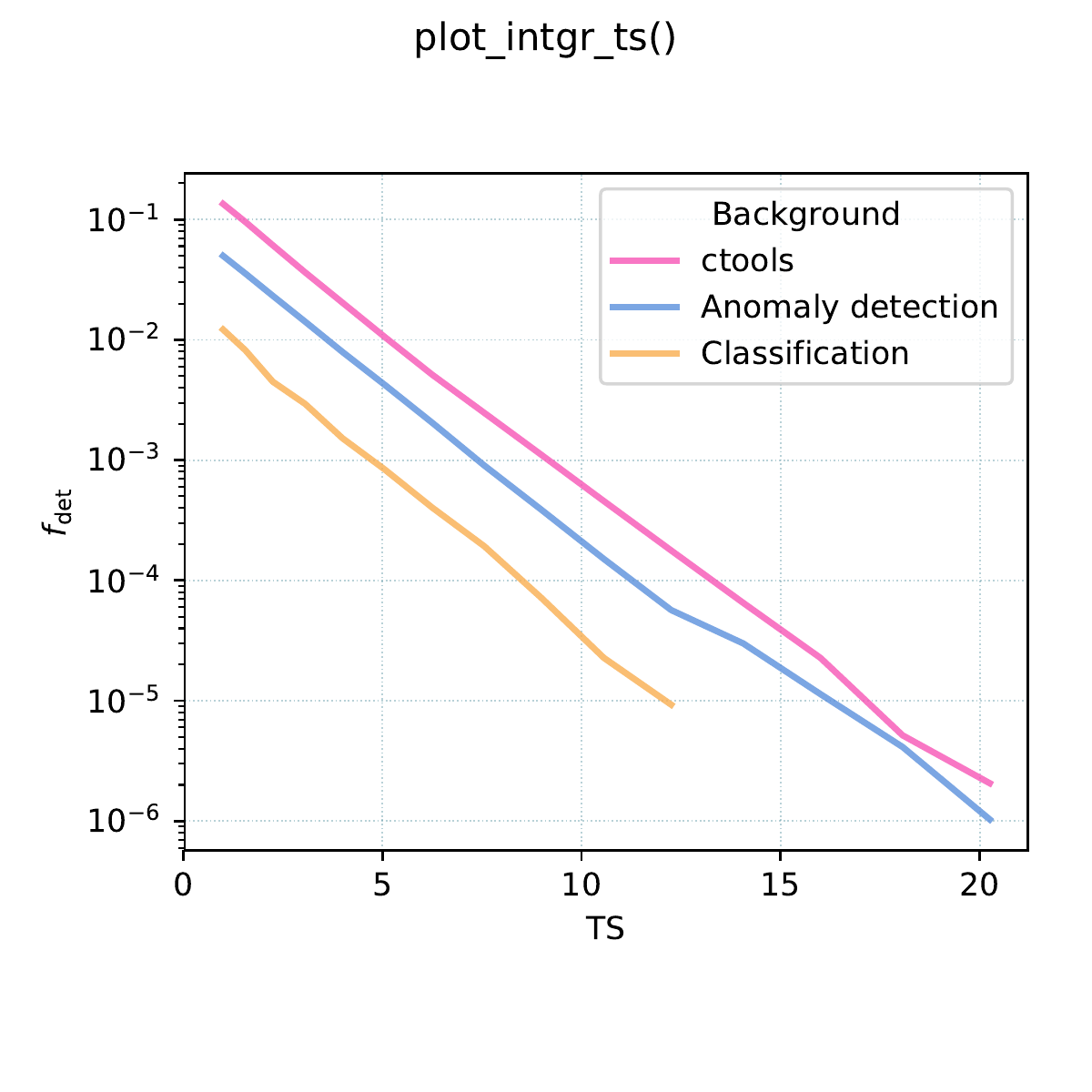}}
            \end{center}\end{minipage}\hfill 
          \end{center}
        \end{minipage}\hfill
        \vspace{5pt}
        \begin{minipage}[c]{1\textwidth}
          \begin{center}
            \begin{minipage}[c]{0.47\textwidth}\begin{center}
              \centering\subfloat[]{\label{FIGintgrTsSigSig}}
            \end{center}\end{minipage}\hfill
            \begin{minipage}[c]{0.47\textwidth}\begin{center}
              \centering\subfloat[]{\label{FIGintgrTsSigBck}}
            \end{center}\end{minipage}\hfill 
          \end{center}
        \end{minipage}\hfill
        %
        %
        \begin{minipage}[c]{1\textwidth}
          \begin{center}
            \begin{minipage}[t]{1\textwidth}\begin{center}
              \caption{\label{FIGintgrTsSig}Dependence
              of \fdet, the fraction of events with a \ts value
              larger than a given threshold,
              on the value of the threshold. The different detections methods
              are compared, derived for the
              signal~\Subref{FIGintgrTsSigSig}
              and background~\Subref{FIGintgrTsSigBck}
              samples, as indicated.
              The full lines in either figure correspond to the
              pre-trials test statistic.
              The dashed lines in~\Subref{FIGintgrTsSigSig}
              represent \fdet after accounting for trials,
              where in~\Subref{FIGintgrTsSigBck}
              we found ${\fdet(\ts>1) = 0}$ post-trials
              in all cases.
              For clarity, the relations are truncated to the
              range, ${1 < \ts < 100}$, where
              the pre-trials background distributions in~\Subref{FIGintgrTsSigBck}
              do not extend beyond ${\ts \approx 20}$.
              The dashed-dotted vertical
              line in~\Subref{FIGintgrTsSigSig} highlights
              the value, $\ts = 25$.}
            \end{center}\end{minipage}\hfill
          \end{center}
        \end{minipage}\hfill
        \vspace{15pt}
      \end{minipage}\hfill

    \end{figure*}

    We now proceed to evaluate the test
    statistics for the various methods.
    \Autoref{FIGintgrTsSig} shows 
    \fdet, the fraction of events with a \ts value
    larger than a given threshold,
    as a function of this threshold.
    One may compare the performance of the
    different detection methods for signal events.

    We find that \ctools
    and the \andet achieve comparable
    significance distributions. The two methods
    manage to detect a similar
    fraction of the events,
    with slightly better performance by \ctools.
    The baseline equivalence between
    the methods is expected, as both utilise
    Poissonian statistics. However, \ctools
    gains significance from a successful fit
    to the assumed spectral model of the source,
    while no such assumptions are taken
    for the \andet. This relative gain in performance
    is balanced out when the intrinsic spectral
    model of the source is more complicated, as discussed below.

    Considering the \cls approach, the performance
    is better than that
    of \ctools, with a relative improvement
    in detectability of~${\Sim10\%}$ on average.
    A direct comparison between these two techniques
    is less straightforward. 
    In general, the assumed temporal properties
    of a transient may be incorporated into
    a likelihood analysis.
    As we did not take this approach in the current study,
    the \rnn has a clear advantage over \ctools.
    In addition, the relative weights between the
    different energy and time bins are
    optimised as a part of training.
    The \cls method is therefore 
    less sensitive to the
    intrinsic spectra of the sources,
    which results in increased sensitivity.

    It is important to verify that the new detection
    methods do not produce spurious detections, and that
    the corresponding test statistics are properly mapped to
    significance. We therefore evaluate the different
    algorithms on the background sample,
    and compare them to the reference \ctools distribution.

    As shown in \autoref{FIGintgrTsSigBck},
    the \andet and \cls methods produce
    comparable or better (lower) rates
    of fake detections. The \cls method in particular
    exhibits an order of magnitude relative improvement
    over \ctools.
    For the given sample of \powB{6} background simulations,
    none of the methods exceed a pre-trials \ts value of~$20$,
    or a post-trials value of~$1$.

    \begin{figure*}[tp]

      \begin{minipage}[c]{1\textwidth}

        \begin{minipage}[c]{1\textwidth}
          \begin{center}
            \begin{minipage}[c]{0.47\textwidth}\begin{center}
              \colorbox{figColBox}{\includegraphics[trim=0mm 12mm 0mm 18mm,clip,width=.98\textwidth]{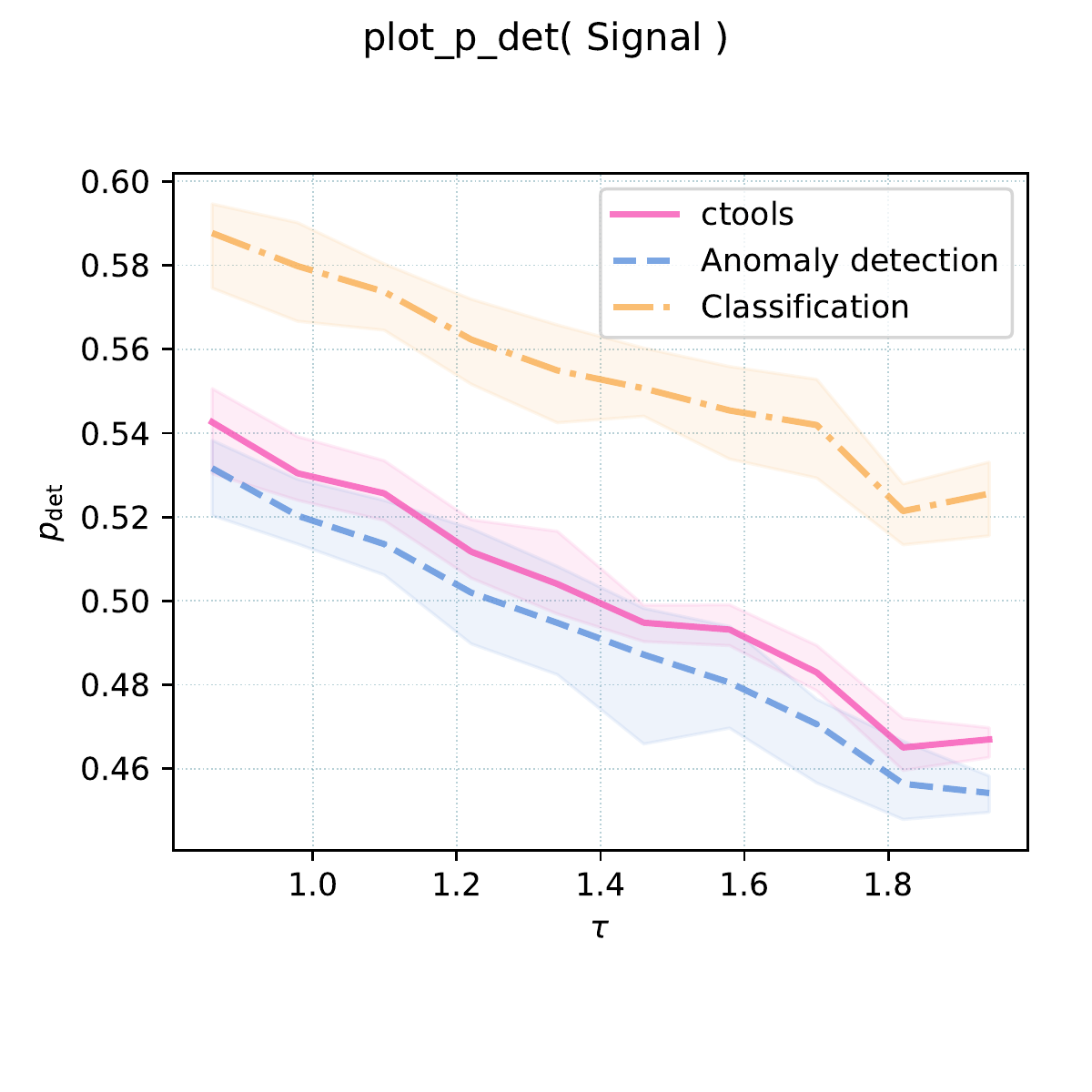}}
            \end{center}\end{minipage}\hfill
            \begin{minipage}[c]{0.47\textwidth}\begin{center}
              \colorbox{figColBox}{\includegraphics[trim=0mm 12mm 0mm 18mm,clip,width=.98\textwidth]{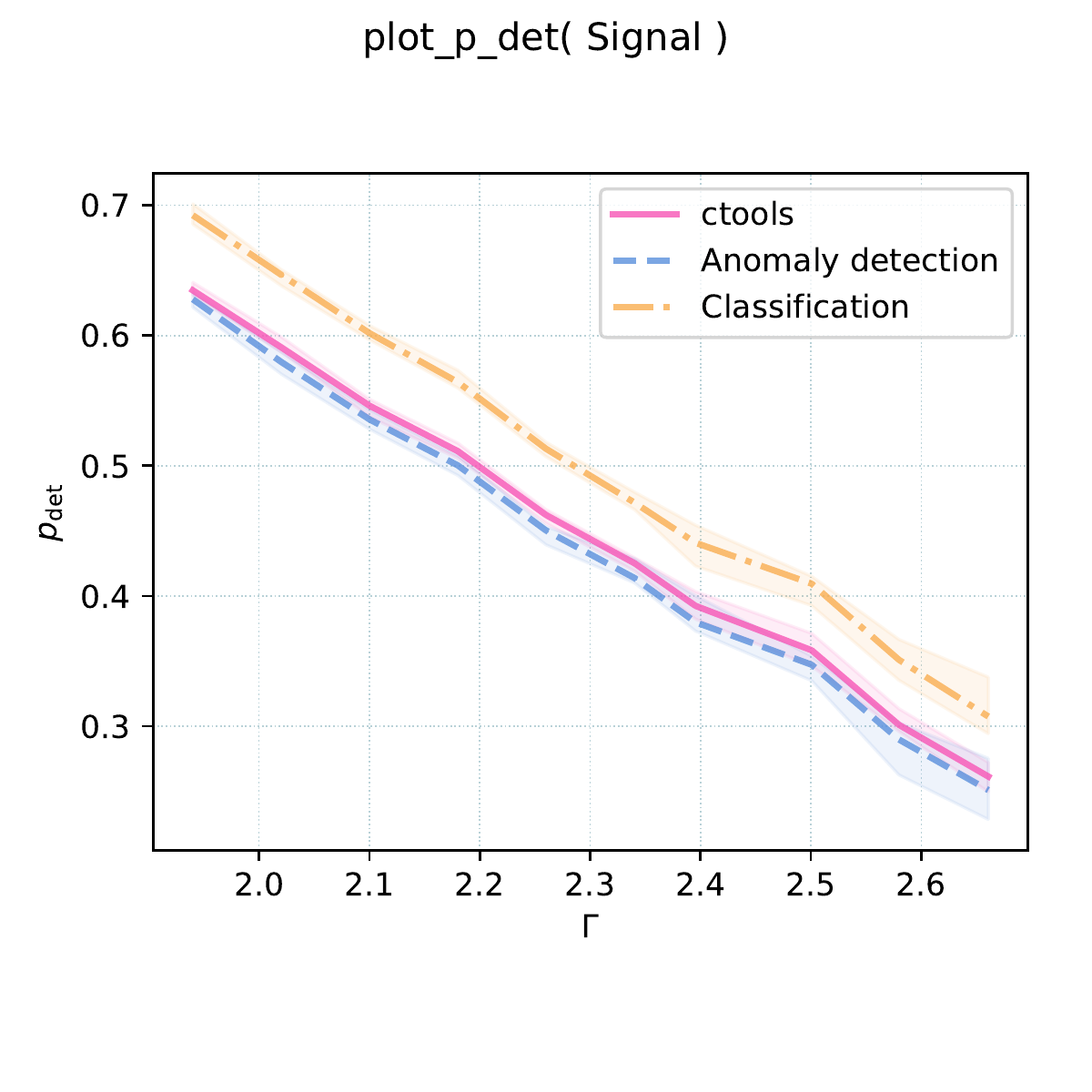}}
            \end{center}\end{minipage}\hfill 
          \end{center}
        \end{minipage}\hfill

        \vspace{-5pt}
        \begin{minipage}[c]{1\textwidth}
          \begin{center}
            \begin{minipage}[c]{0.47\textwidth}\begin{center}
              \centering\subfloat[]{\label{FIGpdetTimeIndex}}
            \end{center}\end{minipage}\hfill
            \begin{minipage}[c]{0.47\textwidth}\begin{center}
              \centering\subfloat[]{\label{FIGpdetSpecIndex}}
            \end{center}\end{minipage}\hfill 
          \end{center}
        \end{minipage}\hfill
        %
        %
        %
        \begin{minipage}[c]{1\textwidth}
          \begin{center}
            \begin{minipage}[c]{0.47\textwidth}
              \begin{center}
                \begin{minipage}[c]{1\textwidth}\begin{center}
                  \colorbox{figColBox}{\includegraphics[trim=0mm 12mm 0mm 18mm,clip,width=.98\textwidth]{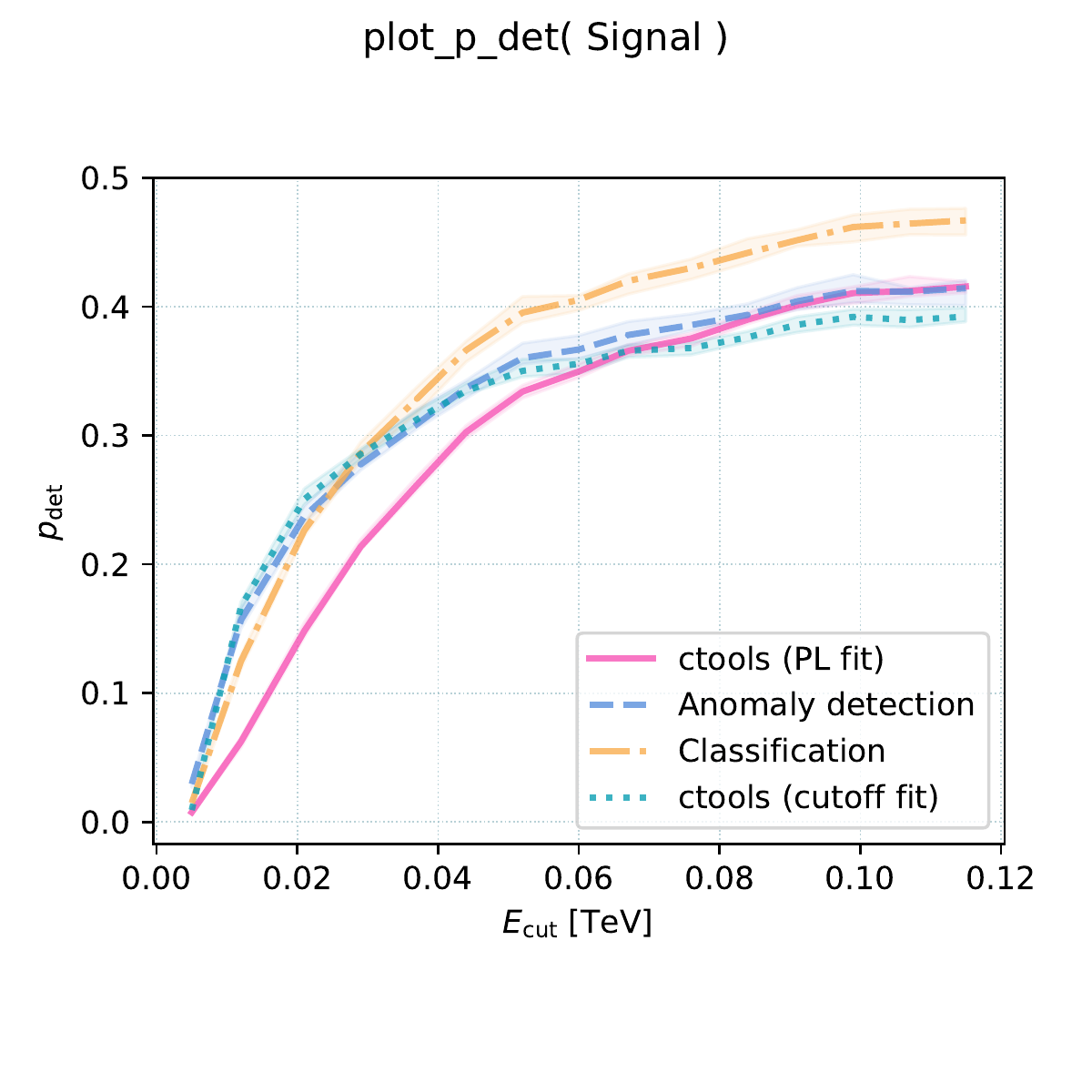}}
                \end{center}\end{minipage}\hfill
                \vspace{-5pt}
                \begin{minipage}[c]{1\textwidth}\begin{center}
                  \centering\subfloat[]{\label{FIGpdetExCutoff}}
                \end{center}\end{minipage}\hfill 
              \end{center}
            \end{minipage}\hfill
            \begin{minipage}[c]{0.47\textwidth}\begin{center}
              \caption{\label{FIGpdetSpecIndexExCutoff}Dependence
              of \pdet on selected simulation parameters,
              after accounting for trials,
              where the shaded regions correspond to~${1\sigma}$
              uncertainties on the values of \pdet, derived
              using the bootstrap method.
              \Subref{FIGpdetTimeIndex}-\Subref{FIGpdetSpecIndex}~Dependence
              of \pdet on the temporal and spectral
              indices of \grbs, $\tau$ and $\Gamma$,
              for bursts simulated as \pls (see \autoref{eqGrbPlModel}),
              for the different detection methods, as indicated.
              \Subref{FIGpdetExCutoff}~Dependence of \pdet
              on the cutoff energy, \ecut,
              for bursts simulated as exponentially cutoff \pls
              (see \autoref{eqGrbExCutoffModel}).
              Two alternative models are used
              for detection with \ctools,
              an exponentially cutoff \pl,
              and a simple \pl, as indicated. The
              \cls method had exclusively
              been trained using simple \pl
              examples in all cases.}
            \end{center}\end{minipage}\hfill 
          \end{center}
        \end{minipage}\hfill
        %
        %
      \end{minipage}\hfill

    \end{figure*} 

    We investigated different combinations of
    \grb spectral and temporal parameters. The most
    important of the latter are those which moderate
    the duration of the burst, and
    the extension of the spectra to multi-\gev energies;
    namely, the temporal and spectral indices, and the
    possible existence of a cutoff.
    The dependence of \pdet
    on these parameters
    is shown in \autoref{FIGpdetSpecIndexExCutoff}.
    One may observe that our new methods
    match or improve upon the performance of \ctools.
    As expected, longer-lasting and harder spectra
    are more likely to be detected by all algorithms.

    \Autoref{FIGpdetExCutoff}
    shows the detectability of \llgrbs
    which are simulated with
    exponentially cutoff \pl spectra.
    For cutoff energies, ${\ecut < 1~\gev}$, bursts
    are undetectable
    within the chosen reconstructed energy range
    (${30 < E_{\gamma} < 200~\gev}$).
    Above this threshold,
    the performance depends on the given detection method.

    The figure illustrates the main motivation
    for using our \rnn. 
    A search with \ctools, under the
    assumption of an exponentially cutoff power law,
    was able to match the performance of the \rnn.
    However, the results of the likelihood analysis
    were not robust (the fits had to be tuned
    with particular choices of parameter initialisation
    and allowed ranges). This is mainly due
    to the relatively low number of \gamrays
    that are available to be fit.

    In practice, the problem of testing 
    complicated models as part of an online
    search is compounded, as
    many possible extensions are possible.
    This implies that one would need
    to make additional assumptions; and
    to test different parameters that
    are not initially well constrained.
    It is therefore doubtful that such a
    search would be successful
    as part of a realistic detection strategy.

    Instead, it is likely that simple \pl models
    will be assumed for the initial blind search.
    Our new algorithms are specifically
    designed to have as little dependence as possible
    on the intrinsic spectra of sources.
    As illustrated here,
    they perform comparably better than
    \ctools in this scenario.


    In principle, the \cls results may be improved,
    by training with
    labelled examples of both \pl and exponentially cutoff
    \grbs. 
    In order to minimise modelling,
    we only used simple
    \pl bursts as examples in the current study,.
    Despite this, the classification is shown to be robust.
    It outperforms the likelihood method (for ${\ecut > 30~\gev}$), even
    when confronted with events representing unexpected
    intrinsic models.

  \section{Summary}
    %
    In this study, we present a new approach for
    source detection.
    Our algorithm is based on deep learning, utilising
    recurrent neural networks, which are ideally suited
    for time series analysis.
    The model can be used to evaluate
    observation sequences
    of second time scales with insignificant latency.
    The choice of technology is therefore
    particularly fitting for real-time searches.

    We have developed two methods, 
    based on \andet and \cls techniques.
    \Andet represents a model-independent
    approach, where transient events are identified based on
    their divergence from the expected background.
    The method is completely data-driven. We thus
    avoid the need for
    background modelling, as well as for
    explicit characterisation of the state of the instrument.

    The \cls method allows one to perform
    targeted searches. In this case, the \rnn
    is trained to identify generic transient patterns.
    The estimator
    provides high detection rates while maintaining
    low fake rates.

    We have compared the performance of
    our new methods to that of
    existing techniques,
    where the background and
    signal models are explicitly defined.
    With the new algorithms,
    we are able to match or improve upon the existing
    methods, while using
    fewer assumptions.
    This is especially important when non-trivial
    source models are considered, \eg
    exponentially cutoff power law spectra. In such cases,
    we have shown that our approach is more robust.

    We have used our new methodology
    to derive the detection prospects of \llgrbs with \cta.
    Provided that \llgrbs indeed exhibit \vhe emission
    above ${\Sim30~\gev}$,
    \cta will be sensitive to a wide range of events.
    Depending on their redshift, bursts with
    isotropic equivalent luminosities as low as
    ${\powB{46}~\erg\,\sec^{-1}}$
    could be detected.

    While we have used \llgrbs as a benchmark
    source class, the methodology presented here
    is applicable for any transient search. Our
    \rnn estimator can trivially be generalised
    for searches over different time scales. It can
    also be used with different types of
    inputs, unrelated
    to photon counts, \eg images and
    other analysis products.
    As such it is ideally suited for
    multiwavelength and multi-messenger transient searches.

  \acknowledgments
    %
    We would like to thank the following
    people for numerous useful discussions:
    D.\,Biehl,
    D.\,Boncioli,
    Z.\,Bosnjak,
    O.\,Gueta,
    T.\,Hassan,
    M.\,Krause,
    G.\,Maier,
    M.\,Nievas Rosillo,
    A.\,Palladino,
    E.\,Pueschel,
    R.\,R.\, Prado,
    L.\,Rauch,
    and
    W.\,Winter.

    This research made use of \ctools, a community-developed analysis package for Imaging Air Cherenkov Telescope data. \ctools is based on \gamlib, a community-developed toolbox for the high-level analysis of astronomical gamma-ray data.

    This research has made use of the \cta \irfs, provided by the \cta Consortium and Observatory, see~\url{http://www.cta-observatory.org/science/cta-performance/} (version \prodIrf) for more details.

    \software{\tensorflow \citep{tensorflow2015-whitepaper};
    \ctools \citep{Knodlseder:2016nnv}}.

  \bibliographystyle{aasjournal}
  \bibliography{bib}

\end{document}